\pdfoutput=1
\documentclass[runningheads]{llncs}

\usepackage{graphicx}
\usepackage{xspace}
\usepackage{multicol}
\usepackage{multirow}
\usepackage{enumitem}
\usepackage{mathtools}
\usepackage{caption}
\usepackage{subcaption}
\usepackage{array}
\newcolumntype{L}[1]{>{\raggedright\let\newline\\\arraybackslash\hspace{0pt}}m{#1}}
\newcolumntype{C}[1]{>{\centering\let\newline\\\arraybackslash\hspace{0pt}}m{#1}}
\newcolumntype{R}[1]{>{\raggedleft\let\newline\\\arraybackslash\hspace{0pt}}m{#1}}
\newcommand{\ethereum}{\hbox{Ethereum}\xspace}

\newcommand{\usdt}{\hbox{\texttt{USDC}}\xspace}
\newcommand{\dai}{\hbox{\texttt{DAI}}\xspace}
\newcommand{\ampl}{\hbox{\texttt{Ampleforth}}\xspace}
\newcommand{\bac}{\hbox{\texttt{Basis Cash}}\xspace}
\newcommand{\frax}{\hbox{\texttt{Frax}}\xspace}
\newcommand{\baccoin}{\hbox{\texttt{BAC}}\xspace}
\newcommand{\bascoin}{\hbox{\texttt{BAS}}\xspace}
\newcommand{\babcoin}{\hbox{\texttt{BAB}}\xspace}
\newcommand{\fraxcoin}{\hbox{\texttt{FRAX}}\xspace}
\newcommand{\fxscoin}{\hbox{\texttt{FXS}}\xspace}
\newcommand{\uniswap}{\hbox{\texttt{Uniswap}}\xspace}
\newcommand{\amm}{\hbox{AMM}\xspace}
\newcommand{\dex}{\hbox{DEX}\xspace}

\newcommand{\data}{\hbox{\texttt{https://explore.duneanalytics.com/dashboard/winky}}\xspace}

\newcommand{\myparagraph}[1]{\vspace*{0.14cm}\noindent\textbf{\emph{#1.}}\quad}

\newcommand{\etal}{\hbox{\emph{et al.}}\xspace}
\newcommand{\eg}{\hbox{\emph{e.g.}}\xspace}
\newcommand{\ie}{\hbox{\emph{i.e.}}\xspace}

\begin{document}
%
\title{Understand Volatility of Algorithmic Stablecoin: Modeling, Verification and Empirical Analysis}
%
%

\author{Wenqi Zhao \and Hui Li \and Yuming Yuan}

\authorrunning{W. Zhao et al.}
%

\institute{
	Huobi Research, Hainan, China\\
	\email{\{zhaowenqi,lihui0729,yuanyuming\}@huobi.com}
}
\maketitle              
\begin{abstract}
An algorithmic stablecoin is a type of cryptocurrency managed by 
algorithms (\ie, smart contracts) to dynamically minimize the volatility of its 
price relative to a specific form of asset, \eg, US dollar. As 
algorithmic stablecoins have been growing rapidly in recent years, they become 
much more volatile than expected. In this paper, we took a deep dive into 
the core of algorithmic stablecoins and shared our answer 
to two fundamental research questions, \ie, Are algorithmic stablecoins volatile by design? 
Are they volatile in practice? Specifically, we introduced an in-depth study on 
three popular types of algorithmic stablecoins and developed a modeling framework to 
formalize their key design protocols. Through formal verification, the framework can 
identify critical conditions under which stablecoins might become volatile. Furthermore, 
we performed a systematic empirical analysis on real transaction activities of the \bac 
stablecoin to relate theoretical possibilities to market observations. Lastly, 
we highlighted key design decisions for future development of algorithmic stablecoins.

\end{abstract}

\section{Introduction}
\label{sec:intro}

As cryptocurrencies on blockchain are notoriously known as volatile, \ie, Their prices 
often fluctuate rapidly, stablecoins are proposed to peg their value to some external assets, 
\eg, US dollar. In contrast to ``unstable'' 
cryptocurrencies, \eg, Bitcoin~\cite{nakamoto2019bitcoin}, \ethereum~\cite{wood2014ethereum}, 
a stablecoin is able to minimize the volatility 
of its price relative to the pegged asset based on different mechanisms. The most 
common kind of stablecoins is backed-stablecoin, \ie, The value of a stablecoin is 
backed by external assets, \eg, commodity, fiat money or cryptocurrency as collateral. 
For example, the \usdt stablecoin is backed by US dollar~\cite{usdt}. 
Unlike backed-stablecoins, algorithmic stablecoins, which are commonly not backed 
by other assets, have been gaining an increasing level of popularity in recent years 
due to the capability to stabilize its price via decentralized algorithms (\ie, smart contract) 
without degrading too much capital efficiency. In general, this is realized by controlling the 
money supply of algorithmic stablecoins, which is similar to printing and destroying 
money in central banks. In this paper, we mainly focus on algorithmic stablecoins and will 
use the term interchangeably with ``stablecoin'' (Backed-stablecoins are not the main target 
in this work.).

Assuming that a stablecoin is pegged to US dollar, a smart contract is designed to 
dynamically manage its supply to minimize price volatility. We simply explain the algorithm 
as follows and will further the discussion later. When the price of the stablecoin 
exceeds one US dollar, the contract ``produces'' more coins and distributes them to 
the market. As a result, the price of the stablecoin should accordingly drop. In cases 
where the price of the stablecoin is lower than one US dollar, the smart contract 
decreases the supply of it in order to gradually lift its price back to one dollar. 
In practice, the aforementioned general algorithm can be instantiated by different 
models to achieve a more robust control over stablecoins. 
While many interesting research attempts aim at inventing such models, there 
is relatively little study on the other side, \ie, Do they really work?

In this paper, we described a fundamental analysis on the volatility of 
algorithmic stablecoins, both theoretically and empirically. Our attempt 
of this study is to answer two fundamental research questions, which are:

\begin{itemize}[leftmargin=39mm]
	
\item[\textbf{Research Question 1:}] Are algorithmic stablecoins volatile \emph{by design}?

\item[\textbf{Research Question 2:}] Are algorithmic stablecoins volatile \emph{in practice}?

\end{itemize}

Our goal of the analysis described in this paper is to provide a more comprehensive 
understanding on the protocols of stablecoins (at both design and implementation level) 
with a specific focus on their volatility, 
which we believe is critical in the optimization of existing stablecoins and creation of 
potential future designs. We summarize our main contributions as follows.

\begin{itemize}[leftmargin=*]
	
\item We introduced an in-depth protocol analysis on the designs of 
three popular types of algorithmic stablecoins. Moreover, we developed a 
general formal modeling and verification framework for stablecoins, which 
can be used to identify specific hidden criteria under which stablecoins might 
become volatile.

\item We further conducted a systematic empirical study of the \bac stablecoin based on 
real transaction activities on \ethereum and manged to relate theoretical 
possibilities (that stablecoins might be volatile) to market observations 
(unexpected volatile prices) between Dec 2020 to Jan 2021.
	
\end{itemize}

\section{Background}
\label{sec:bg}

We classify algorithmic stablecoins into three categories, \ie, rebase-style, 
seigniorage share and partial-collateral. In this section, we briefly explain 
key designs of all three types of stablecoins with popular projects as examples.

\subsection{Rebase (\ampl)}
\label{subsec:rebase}
The rebase-style stablecoins manage price-elastic ERC20 tokens, \ie, The total 
supply of a stablecoin is non-fixed and adjusted adaptively on a routine basis. 
More specifically, the adjustment is automatically realized via the ``rebase'' 
process, which gradually stabilize the price of a target stablecoin near a 
specific peg, \eg, one US dollar. We use \ampl~\cite{ampl} as an example 
for illustration.

By design, the rebasing of \ampl is activated at 2am UTC on a daily basis. 
At the time of rebase, new coins are minted and distributed to all accounts 
proportionally based on their corresponding balances when the price of \ampl 
is higher than its peg. Given that the price of \ampl is \$1.2 with its peg 
to be \$1 (\ie, 20\% relate to peg), an account with 100 coins is rebased to 
own 120. On the other hand, holding coins might be automatically 
proportionally burned when the price falls below the peg.

\subsection{Seigniorage Share (\bac)}
\label{subsec:seigiorage}

The seigniorage share model for algorithmic stablecoins commonly introduces 
two types of cryptocurrencies, \ie, \emph{coins} as a stablecoin and \emph{shares} 
as ownership of seigniorage. In principle, shares are used to increase the supply 
of coins when the price of a coin is above its intended peg. 
In addition to these two cryptocurrencies, seigniorage-style stablecoins often 
issue a redeemable bond as an incentive for buyers when the price 
goes down below the peg. We use the \bac~\cite{bac} 
stablecoin for further explanation. 
\bac introduces three types of cryptocurrencies:

\begin{itemize}[leftmargin=*]
\item \texttt{BAC}. \baccoin is the stablecoin and issued by the \bac 
with a peg of \$1. 

\item \texttt{BAS}. \bascoin stands for Basis Shares, which is a seigniorage ERC20 
token and provides inflationary gains of \baccoin. The design purpose of \bascoin 
is to prevent the price of \baccoin from going too high 
via dynamically increasing its supply. Currently, \bascoin can be 
earned via participating in yield farming, \ie, deposit liquidity in decentralized 
finance platforms (\eg, Uniswap~\cite{uniswap}).

\item \texttt{BAB}. \babcoin refers to Basis Bond whose price $P_{\mathit{bab}}$ 
is mathematically determined by the price of \baccoin $P_{\mathit{bac}}$, 
\ie, $P_{\mathit{bab}} = (P_{\mathit{bac}})^2$. Particularly, \babcoin offers an 
incentive for holders to earn \babcoin in a cost-effective way. The design purpose 
behind is to push \baccoin back to one dollar when its price falls below \$1. 

\end{itemize}

The general protocol of \bac is designed to stabilize the price of \baccoin via adaptively 
controlling the supply of it. This is realized based on the two key mechanisms, \ie, \emph{expansion} 
and \emph{contraction}, respectively. We simply describe the processes as below.

\myparagraph{Expansion}
The mechanism of expansion aims at increasing the supply of \baccoin in order to stabilize its 
price when it rises over the one dollar peg. In the design of \bac, expansion is automatically 
activated in two settings. First, \baccoin will be minted and distributed as a reward 
to \bascoin holders. That said, for anyone who owns a specific amount of \bascoin, the 
expansion process proportionally assigns new \baccoin to his or her account. In the second 
case, owners of \babcoin are allowed to redeem \baccoin with their \babcoin at a 1:1 price, 
which also result in a quantity growth of \baccoin. Due to the increased supply in both 
situations, the expansion is expected to gradually make the price of \baccoin to decrease.

\myparagraph{Contraction}
In contrast to the process of expansion, contraction is designed to shrink the supply of 
\baccoin. To this end, an incentive is introduced in \bac to encourage buyers to exchange 
\babcoin with \baccoin when the price of \baccoin is below one dollar. In the particular 
situation, one \baccoin is guaranteed to generate more than one \babcoin based on their 
price dependency as aforementioned. Moreover, the protocol of \bac ensures that a specific 
amount of \babcoin is able to redeem the same amount of \baccoin when the price of \baccoin 
grows above \$1 and required conditions are met. Based on the design of contraction, th price 
of \baccoin is anticipated not to fall too far from its peg.

\subsection{Partial-Collateral (\frax)}
\label{subsec:partial}
In contrast to the two types of algorithmic stablecoins, an emerging class called 
fractional-algorithmic protocol is recently proposed as a combination of 
fully-collateral and fully-algorithmic ones. Compared to existing collateral-style 
stablecoins, \eg, \dai, partial-collateral protocols introduce less custodial risks 
and avoid over-collateralization. On the other hand, it is designed to enforce a 
relatively tight peg with higher level of stability than purely algorithmic designs. 
We use the \frax project~\cite{frax} below for illustration. 

Particularly, \frax is the first attempt to implement the partial-collateral protocol 
of stablecoins. It introduces a two-token system, \ie, \fraxcoin as a stablecoin pegged 
to \$1 and \fxscoin as a governance token, respectively. A collateral ratio $0\le r \le 1$ is 
dynamically determined very hour with a step of 0.25\% in the protocol to control at what 
percentage of peg the collateral is required to take to stabilize the value of \fraxcoin. 
In cases where $r=0.5$, \$0.5 must be in other types of stablecoins as collateral to 
mint a new \fraxcoin. It becomes fully collateral when $r=1.0$ and a pure 
algorithmic stablecoin if $r=0$. 

The collateral ratio $r$ is 1.0 at genesis. In principle, minting a specific amount $n$ of \fraxcoin 
involves placing $n \times r$ of the value as collateral and burning $n \times (1-r)$ of the value 
with \fxscoin. As the price goes above its peg, the protocol provides the incentive 
for investors to mint new \fraxcoin. Accordingly, the increased 
supply of \fraxcoin is expected to gradually enforce the price to decrease. In cases 
where the price falls below the peg, the protocol allows investors to swap a combination of 
collateral and \fxscoin valued \$1 with a single \fraxcoin whose value is lower than \$1. 
Such incentives can potentially produce \fraxcoin purchases and rise its price as well.

\section{Modeling and Verification}
\label{sec:model}

\subsection{Modeling of Stablecoin}
\label{subsec:modeling}
We highlighted a formal modeling framework $\mathcal{M}$ for stablecoins. More 
formally, $\mathcal{M} \coloneqq \langle \mathcal{P}, \mathcal{E}, \mathcal{C}, 
\mathcal{S}, \mathcal{B}, \mathcal{X}\rangle$ is a network consisting 
of six types of timed automata~\cite{alur1994theory}, each of which is a 
tuple $Q \coloneqq \langle S, s_0, X, A, T, I, S_n \rangle$. $S$ is the finite set of states. 
$s_0 \in S$ is the initial state. $X$ is a set of non-negative real numbers as 
clock variables. $S_n \subseteq S$ is a set of accepting states. $A$ is a set of 
actions and $I$ is a set of invariants assigned to states. Given that $\Phi$ is constraint 
function, 
$T \subseteq S \times \Phi(X) \times 2^X \times A \times S$ is a collection 
of state transitions $\langle s, a, g, R, s' \rangle$, where $s$ and $s'$ 
are source and destination states, $a$ is an action, $g$ is the condition to 
enable the transition and $R$ is the set of clocks to be reset. 

\begin{figure}
	\centering
	\begin{subfigure}[b]{0.3\textwidth}
		\centering
		\includegraphics[width=\textwidth]{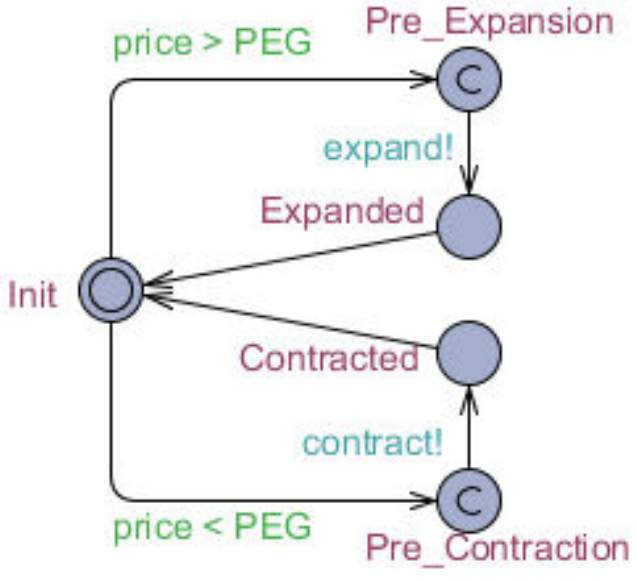}
		\caption{Protocol $\mathcal{P}$}
		\label{fig:protocol}
	\end{subfigure}
	\hfill
	\begin{subfigure}[b]{0.34\textwidth}
		\centering
		\includegraphics[width=\textwidth]{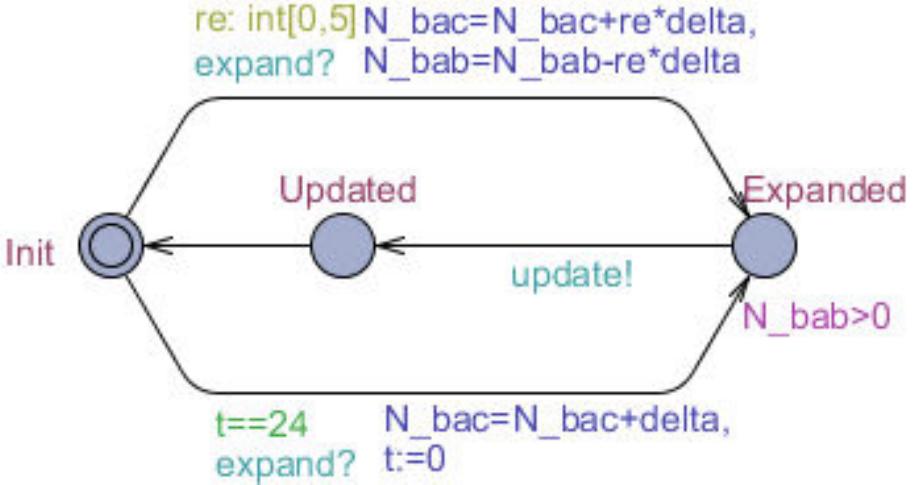}
		\caption{Expansion $\mathcal{E}$}
		\label{fig:expansion}
	\end{subfigure}
	\hfill
	\begin{subfigure}[b]{0.34\textwidth}
		\centering
		\includegraphics[width=\textwidth]{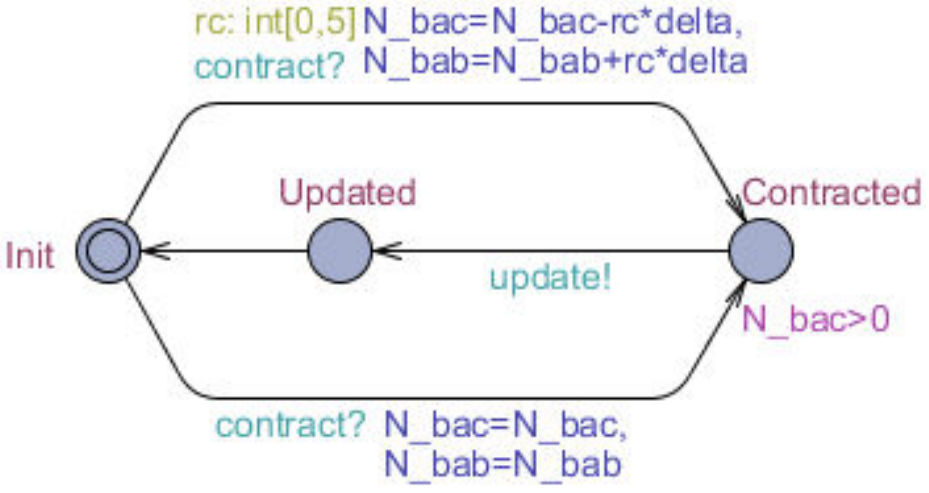}
		\caption{Contraction $\mathcal{C}$}
		\label{fig:contraction}
	\end{subfigure}
	\hfill
	\begin{subfigure}[b]{0.32\textwidth}
		\centering
		\includegraphics[width=\textwidth]{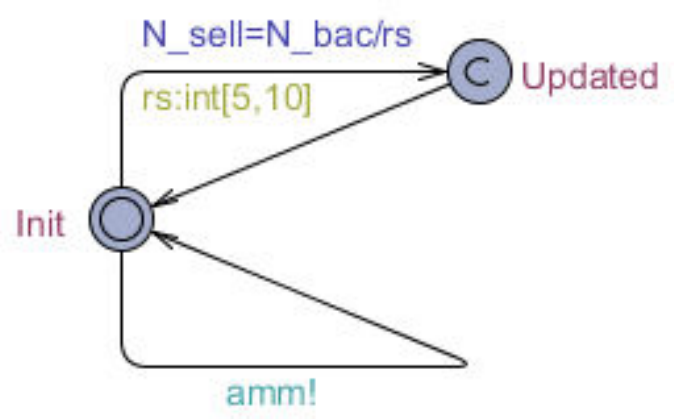}
		\caption{Seller $\mathcal{S}$}
		\label{fig:supply}
	\end{subfigure}
	\hfill
	\begin{subfigure}[b]{0.32\textwidth}
		\centering
		\includegraphics[width=\textwidth]{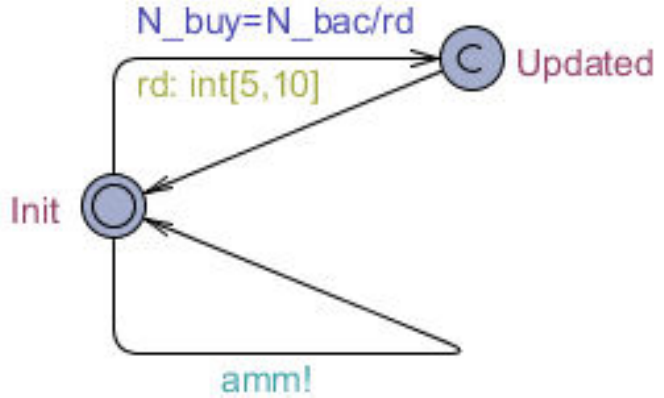}
		\caption{Buyer $\mathcal{B}$}
		\label{fig:demand}
	\end{subfigure}
	\hfill
	\begin{subfigure}[b]{0.34\textwidth}
		\centering
		\includegraphics[width=\textwidth]{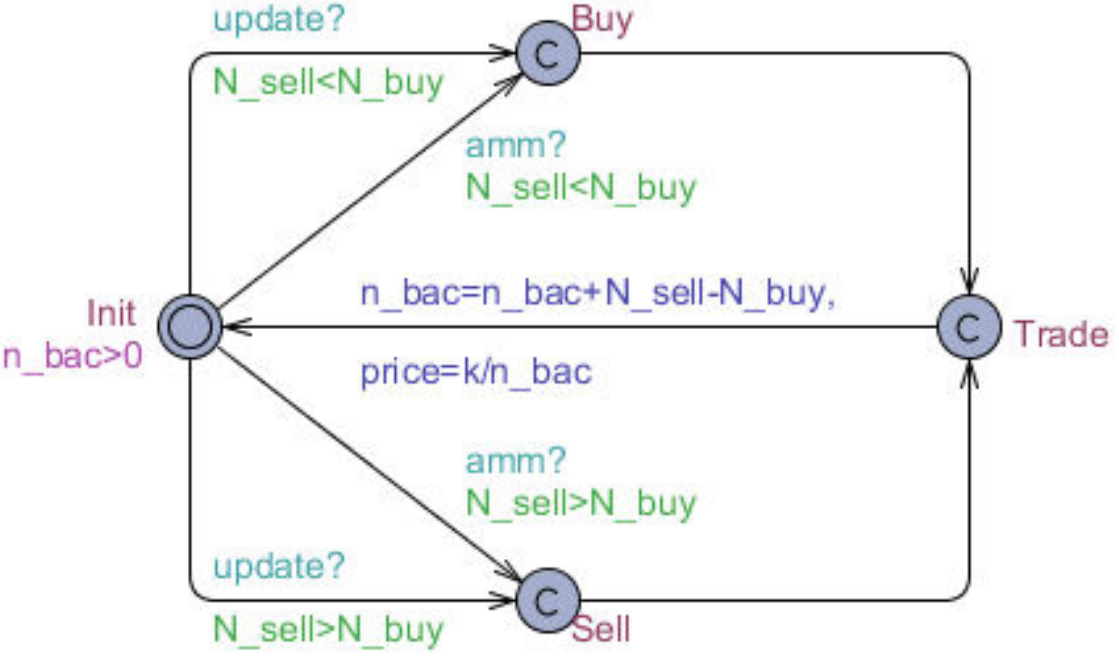}
		\caption{\dex $\mathcal{X}$}
		\label{fig:dex}
	\end{subfigure}
	\caption{Timed automata model of \bac stablecoin.}
	\label{fig:model_bac}
\end{figure}

Moreover, $\mathcal{M}$ provides communication through four classes 
of synchronized channels $\Omega \coloneqq \{ \omega_e, \omega_c, \omega_x, \omega_u \}$. 
Specifically, $\omega_e$ and $\omega_c$ are designed to trigger \emph{expansion} 
and \emph{contraction} procedures. $\omega_x$ simulates 
market trading activities and generates a new price of stablecoin. 
$\omega_u$ synchronizes updates between $\mathcal{E}$, $\mathcal{C}$ and  $\mathcal{X}$. 
Particularly, we presented a formal model of \bac in 
Figure~\ref{fig:model_bac}. The framework is general to other types of stablecoins. 
Due to page limits, we selected \bac because it manifests a typical model and was one 
of the most popular markets at the time of writing.

\begin{itemize}[leftmargin=*]
\item $\mathcal{P}$ models the main protocol with five states, \ie, initial 
state, \texttt{Pre\_Expansion} and \texttt{Expanded} states when price is above the peg, 
\texttt{Pre\_Contraction} and \texttt{Contracted} states when price is below the peg. 
The channels of \texttt{expand} ($\omega_e$) and \texttt{contract} ($\omega_c$) are 
activated on two transitions to enable the processes of expansion and contraction.

\item $\mathcal{E}$ automata defines a process with a clock $t$ and a 
three states. $\mathcal{E}$ responds to expansion requests from $\mathcal{P}$. 
An expanding transition is executed to grow the supply of 
stablecoins (\ie, global variable \texttt{N\_bac}). The transition is allowed if $t$ 
is at an expansion point (\eg, 24:00 UTC). For \bac, $\mathcal{E}$ creates 
two expansion transitions and synchronizes with $\mathcal{X}$ via the \texttt{update} 
channel ($\omega_u$). 

\item $\mathcal{C}$ automata abstracts the contraction process. Similar to 
$\mathcal{E}$, a transition is provided to refine the decrease of supply 
via updating a global variable Another transition is designed to model that 
the supply stays unchanged (investors can choose not to swap \babcoin with \baccoin).  

\item $\mathcal{S}$ and $\mathcal{B}$ are designed to model the behavior of 
sellers and buyers in an exchange. They generate random trading 
requests through the $\omega_x$ channel.  

\item $\mathcal{X}$ introduces an abstract model of decentralized exchanges (\dex) 
with automatic market making (\amm), \eg, \uniswap~\cite{uniswap}. $\mathcal{X}$ 
defines \texttt{Sell} and \texttt{Buy} states to indicate whether it is a buyer's 
market (\ie, more sellers than buyers) or seller's market (the other way around). 
New prices are computed based on \amm and its pool of stablecoins.

\end{itemize}

\subsection{Formal Verification}
\label{subsec:verification}
We further highlight important formal specifications 
to define stability properties (or non-volatility) of 
stablecoins with temporal logic~\cite{pnueli1977temporal}. 
Specifically, stability (non-volatility) is specified through 
the following two properties ($\mathtt{A}$ and $\mathtt{G}$ are quantifiers, \ie, 
for all paths and for all states of a path in the state space~\cite{pnueli1977temporal}).
\begin{equation}\label{eqa:e}
	\mathtt{AG} \ (\mathcal{P}.\mathtt{Expanded} \wedge \mathcal{E}.\mathtt{Validated}) \implies  !\mathcal{X}.\mathtt{Buy}\tag{expansion-validity}
\end{equation}
\begin{equation}\label{eqa:c}
	\mathtt{AG} \ (\mathcal{P}.\mathtt{Contracted} \wedge \mathcal{C}.\mathtt{Validated}) \implies !\mathcal{X}.\mathtt{Sell}\tag{contraction-validity}
\end{equation}

\begin{figure}
	\centering
	\begin{subfigure}[b]{0.45\textwidth}
		\centering
		\includegraphics[width=\textwidth]{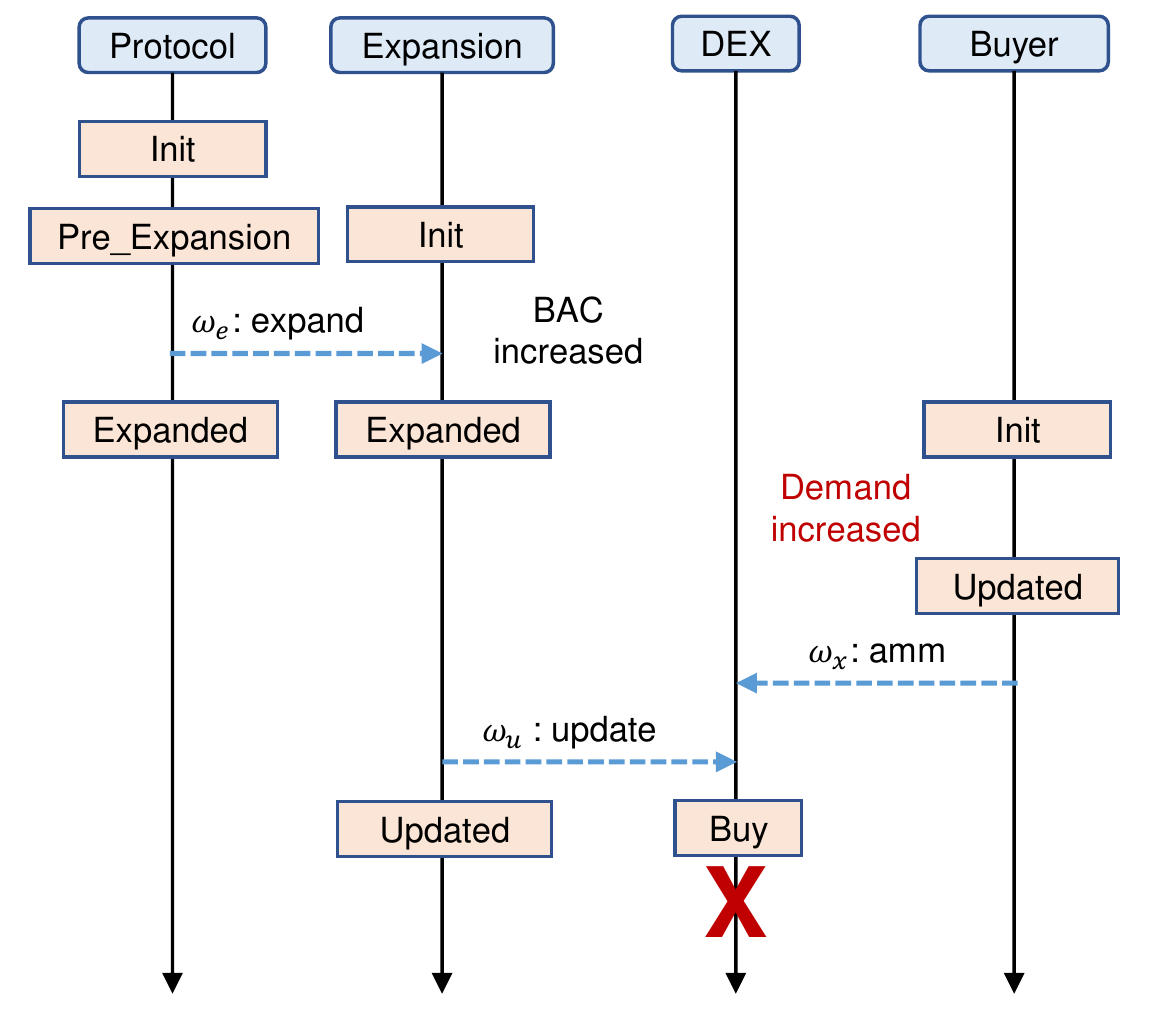}
		\caption{Expansion validity}
		\label{fig:ce_expansion}
	\end{subfigure}
	\hfill
	\begin{subfigure}[b]{0.45\textwidth}
		\centering
		\includegraphics[width=\textwidth]{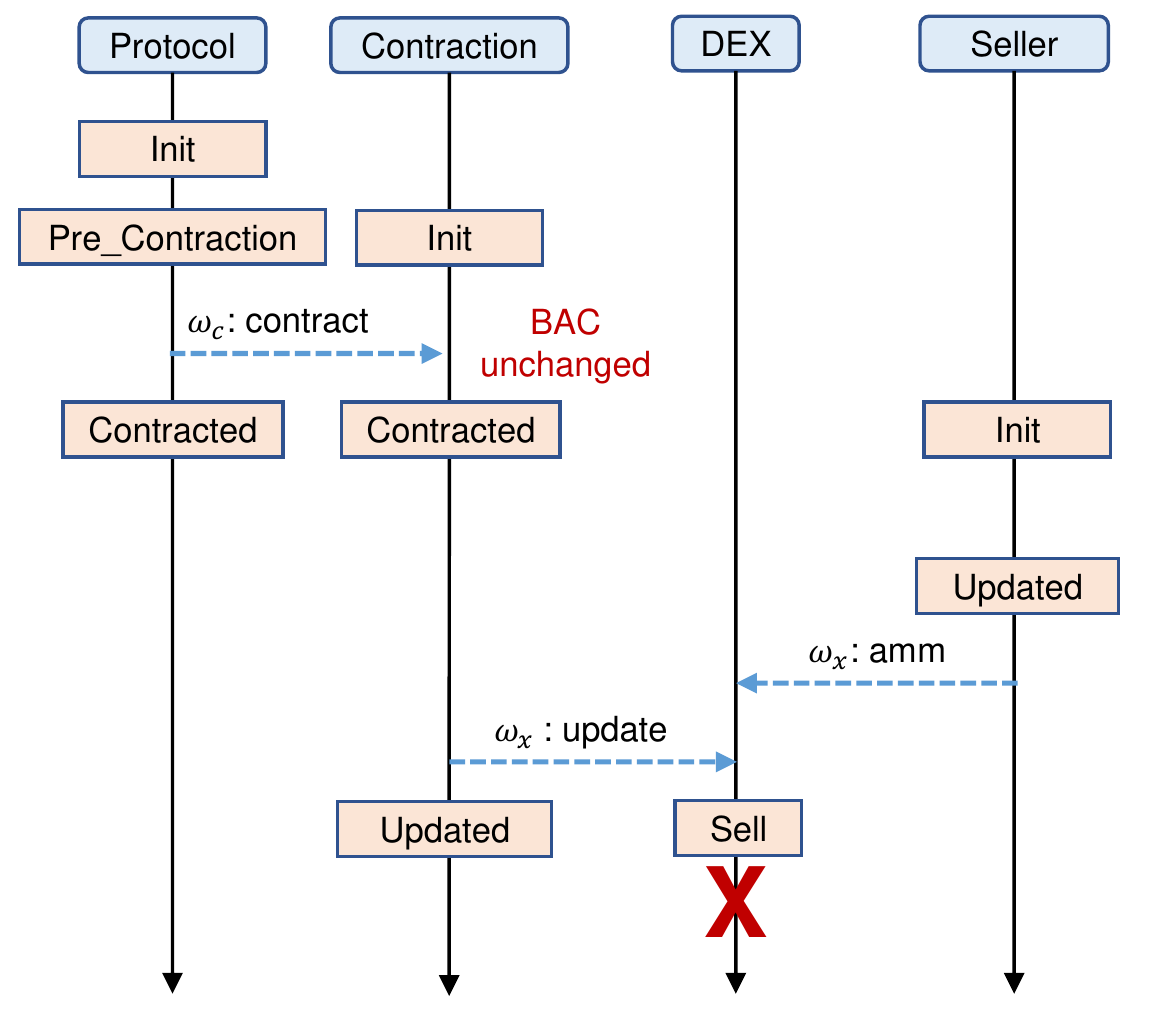}
		\caption{Contraction validity}
		\label{fig:ce_contraction}
	\end{subfigure}
	\caption{Counter-examples on non-volatility properties of \bac stablecoin.}
	\label{fig:ce_bac}
\end{figure}

\myparagraph{Specifications of Stability (Non-Volatility)}
Two properties are specified as in \ref{eqa:e} and \ref{eqa:c} to formalize the 
resilience against price fluctuation (with \bac as an example). 
As formalized by \ref{eqa:e}, in cases where $\mathcal{P}$ 
is at state \texttt{Expansion} and $\mathcal{E}$ is at \texttt{Validated} (\ie, 
expansion has been enforced), $\mathcal{X}$ must 
not stay at the state of \texttt{Buy} for the price to fall, \ie, buyer's market. 
Similarly, when $\mathcal{P}$ is at \texttt{Contraction} and $\mathcal{C}$ is 
at \texttt{Validated}, $\mathcal{X}$ must not be at the state of \texttt{Sell}, 
\ie, seller's market.

\myparagraph{Counter-Example Analysis}
We verified the model of \bac 
with the \texttt{Uppaal} model checker for timed automata~\cite{larsen1997uppaal}. 
Figure~\ref{fig:ce_bac} shows two counter-examples of the stability properties, \ie, 
conditions under which \bac might become volatile. Figure~\ref{fig:ce_expansion} 
describes a trading scenario where expansion validity is violated. Specifically, a 
demand growth of \baccoin occurs when the expansion process is started to mint 
and distribute new stablecoins. As a result, 
\dex goes to the state of \texttt{Buy} instead of \texttt{Sell} to trigger a counter-example. 
In terms of \ref{eqa:c}, Figure~\ref{fig:ce_contraction} demonstrates another 
potential volatility of \bac. When the price of \baccoin goes down below its peg, the 
contraction allows investors to swap \babcoin with \baccoin. However, 
in cases where the swap does not happen therefore supply of \baccoin stays unchanged, 
the \ref{eqa:c} is violated since \dex goes to the state of \texttt{Sell} instead 
of \texttt{Buy} as expected.

\section{Empirical Analysis}
\label{sec:analysis}

Based on the formal modeling and verification of \bac, we now describe an 
empirical analysis with real market observations. Queries and data used are 
available (\data) on the \texttt{Dune Analytics}~\cite{dune} platform.

\begin{figure}
	\centering
	\begin{subfigure}[b]{0.48\textwidth}
		\centering
		\includegraphics[width=\textwidth]{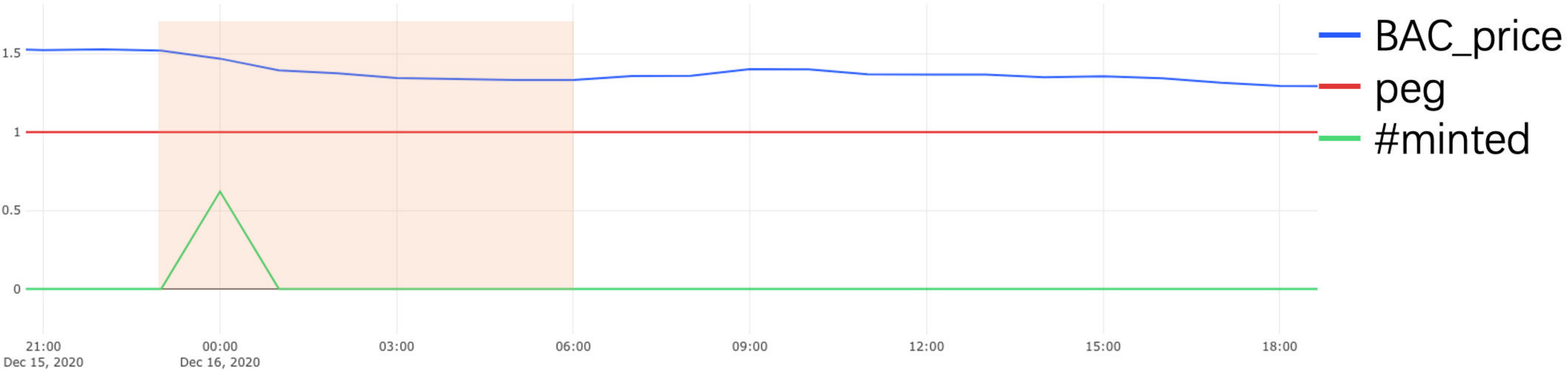}
		\caption{Effective expansion}
		\label{fig:normal_expansion}
	\end{subfigure}
	\hfill
	\begin{subfigure}[b]{0.48\textwidth}
		\centering
		\includegraphics[width=\textwidth]{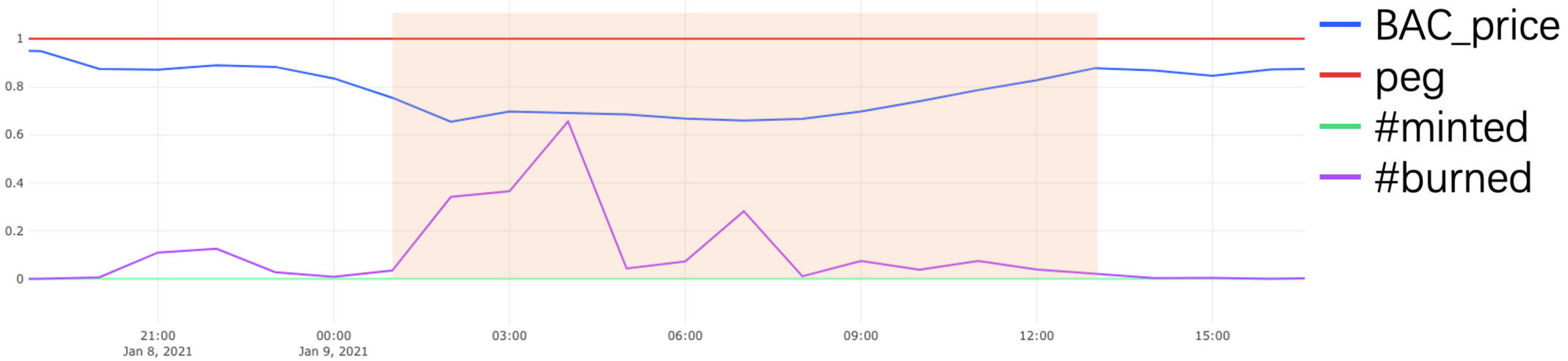}
		\caption{Effective contraction}
		\label{fig:normal_contraction}
	\end{subfigure}
	\hfill
	\begin{subfigure}[b]{0.48\textwidth}
		\centering
		\includegraphics[width=\textwidth]{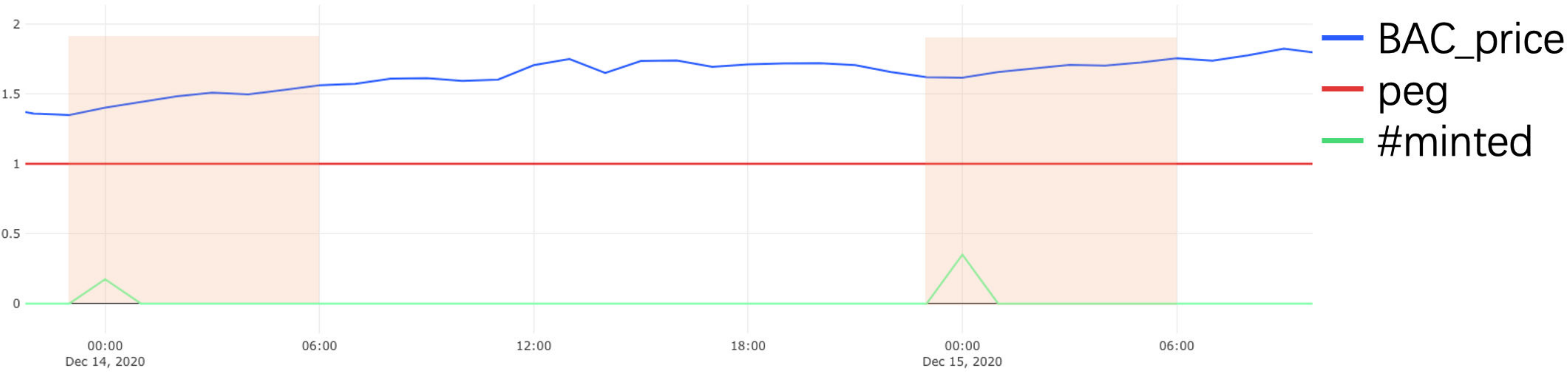}
		\caption{Broken expansion}
		\label{fig:abnormal_expansion}
	\end{subfigure}
	\hfill
	\begin{subfigure}[b]{0.48\textwidth}
		\centering
		\includegraphics[width=\textwidth]{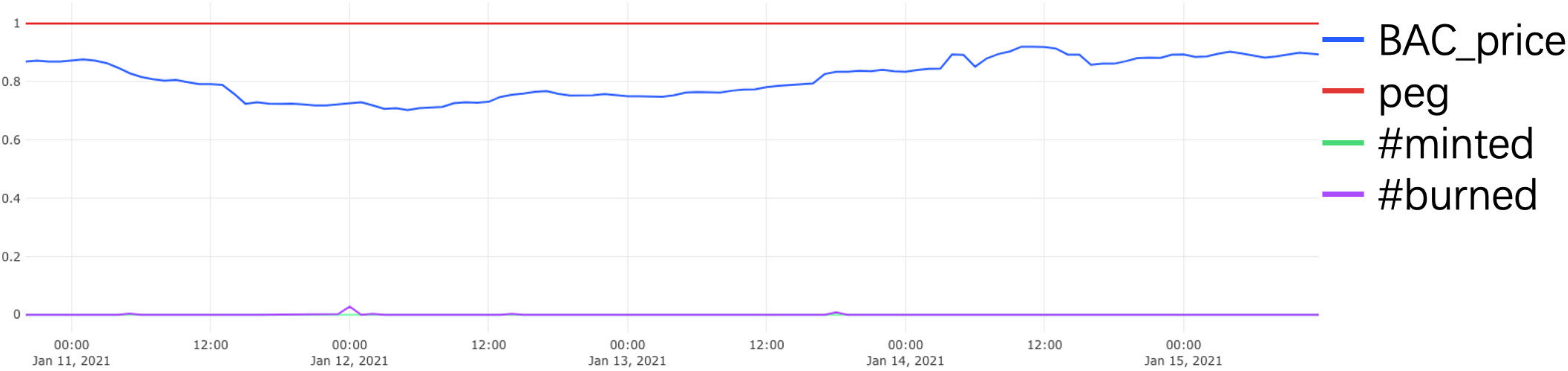}
		\caption{Broken contraction}
		\label{fig:abnormal_contraction}
	\end{subfigure}
	\hfill
	\begin{subfigure}[b]{0.48\textwidth}
		\centering
		\includegraphics[width=\textwidth]{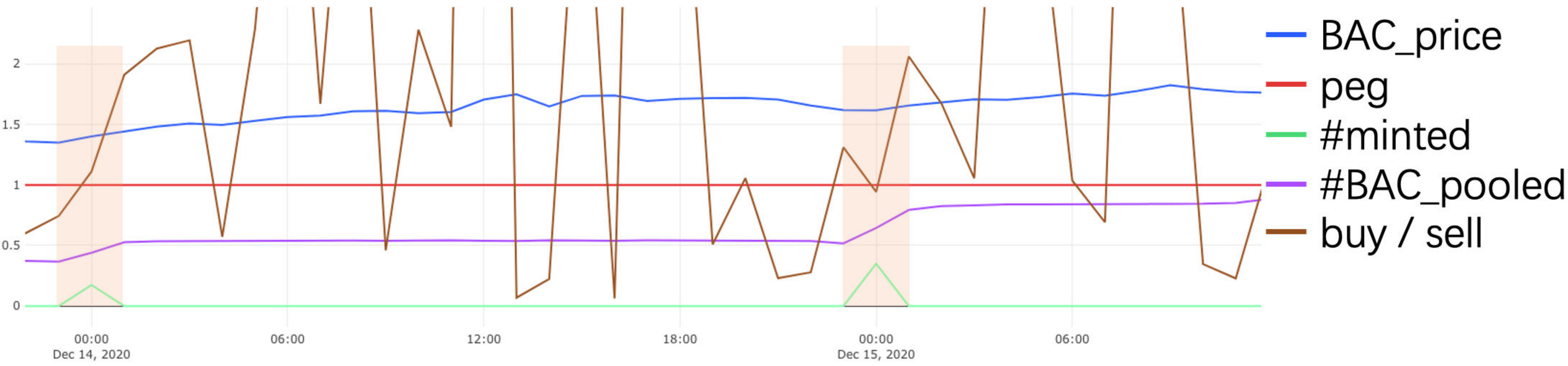}
		\caption{Cause of broken expansion}
		\label{fig:abnormal_expansion_cause}
	\end{subfigure}
	\hfill
	\begin{subfigure}[b]{0.48\textwidth}
		\centering
		\includegraphics[width=\textwidth]{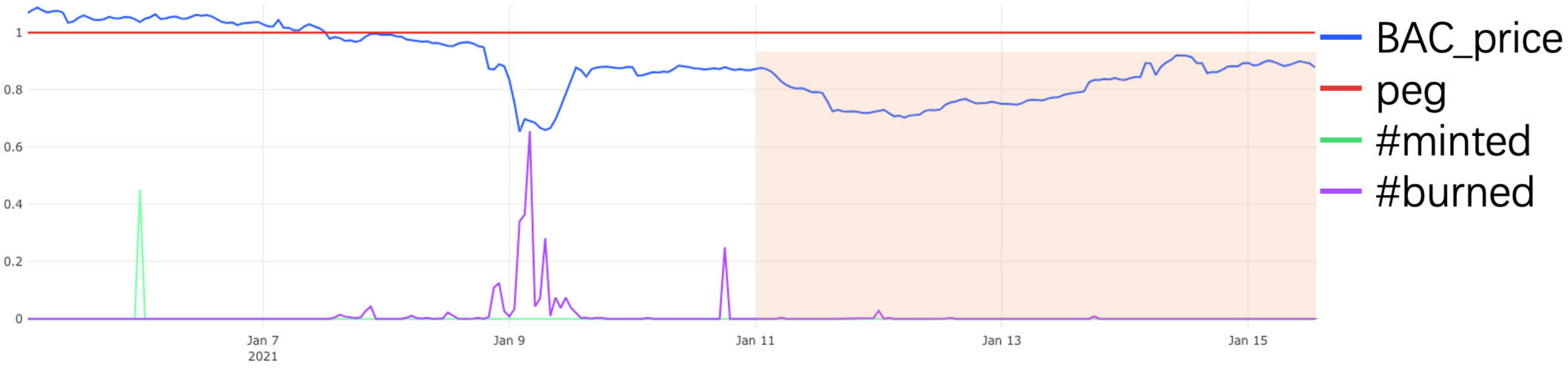}
		\caption{Cause of broken contraction}
		\label{fig:abnormal_contraction_cause}
	\end{subfigure}
	\caption{Empirical analysis of \bac. Unit: 10 million.}
	\label{fig:empirical_bac}
\end{figure}

\myparagraph{Normal Cases}
Figure~\ref{fig:normal_expansion} and \ref{fig:normal_contraction} shows two 
cases where expansion and contraction worked well. 
As highlighted in Figure~\ref{fig:normal_expansion}, as new 
\baccoin were minted, its price gradually went down. 
Similarly, as a number of \baccoin were burned for contraction in 
Figure~\ref{fig:normal_contraction}, its price started to rise.

\myparagraph{Broken Expansion}
Figure~\ref{fig:abnormal_expansion} and \ref{fig:abnormal_expansion_cause} 
explains a broken expansion as inferred in \S\ref{subsec:verification} on 
Dec 14 and 15, 2020. In Figure~\ref{fig:abnormal_expansion}, the expansion 
started at 00:00 with a collection of new \baccoin minted. However, its 
price increased in 7 hours from \$1.35 to \$1.56 (Dec 14) and from \$1.62 to \$1.76 
(Dec 15), which amounted to a growth of 15.72\% and 8.40\%. 
Based on Figure~\ref{fig:abnormal_expansion_cause}, the broken 
expansion was attributed to a rapid increase of demand as marked in 
Figure~\ref{fig:abnormal_expansion_cause}. Since yield-farming on 
\baccoin-\dai was very popular at an early stage and led to a extremely 
high yield rate, the demand of \baccoin was rapidly lifted even 
at an expansion point. The popularity of \baccoin was also 
reflected by that 92\% of the newly minted \baccoin on Dec 14, 2020 went 
to the yield-farming pool within 2 hours after expansion.

\myparagraph{Broken Contraction}
The potential volatility due to broken contraction was also confirmed in 
Figure~\ref{fig:abnormal_contraction} and \ref{fig:abnormal_contraction_cause}. 
From Jan 11, 2021 to the time of writing, the price of \baccoin has been 
staying below its peg according to Figure~\ref{fig:abnormal_contraction} 
despite that entries of contraction were continuously open. The reason 
behind was the low participation in contraction, \ie, Many investors were 
unwilling to burn \baccoin for \babcoin due to the fear that they might never 
be able to redeem. As shown in Figure~\ref{fig:abnormal_contraction_cause}, 
the number of burned \baccoin (hump) in that period was much smaller than 
several days ago.

\myparagraph{Design Decisions}
First, contraction weighs more than expansion to minimize volatility since 
cryptocurrencies are naturally easier to fall than rise. Interfacing contraction 
to more ecosystems is a vital complement. Second, the quantity 
and cycle of algorithmic intervention are essential factors in stablecoin designs. 
More robust and flexible models are highly desired in this context.

\section{Related Work}
\label{sec:rw}

As stablecoins became popular in recent years, researchers have been 
suggesting to a new monetary policy for them~\cite{iwamura2019can,caginalp2018opinion}. 
Saito~\etal proposed to stabilize cryptocurrencies via automatically controlling their supply to 
absorb both positive and negative demand shocks~\cite{saito2019make}. 
Caginalp~\etal leveraged asset flow equations to model cryptocurrency and 
their stability~\cite{caginalp2018dynamical}. 
In the context of algorithmic stablecoins, Ametrano described \emph{Hayek Money} 
to achieve stability via rebasing the amount of coins~\cite{ametrano2016hayek}. 
Sams further designed \emph{seigniorage shares} to include an elastic supply rule 
which adjusts the quantity of coins adaptively~\cite{sams2015note}. On the other 
hand, design review of stablecoins were al discussed in several research 
papers and industry reports~\cite{clark2019sok,moin2020sok,mita2019stablecoin,pernice2019monetary,bullmann2019search,hileman2019state,klages2020stablecoins}. 
Classification of stablecoins were introduced according to different 
types of collateral and intervention with pros and cons explained.

\section{Conclusion}
\label{sec:conclusion}
In this paper, we presented an in-depth theoretical and empirical 
analysis on the volatility of algorithmic stablecoins. We highlighted 
a formal modeling framework for stablecoins to identified important 
market criteria under which they might become volatile. Moreover, 
we related our theoretical findings to transaction activities 
on stablecoins via a further empirical analysis with real market 
data. Empirical results showed that potential possibilities predicted 
in the proposed model were confirmed in practice. 
Lastly, we highlighted important design decisions for the future 
development of stablecoin. All data used in this work are available 
at \data.

%
%
%
\bibliographystyle{splncs04}
\bibliography{stablecoin}

\begin{thebibliography}{10}
\providecommand{\url}[1]{\texttt{#1}}
\providecommand{\urlprefix}{URL }
\providecommand{\doi}[1]{https://doi.org/#1}

\bibitem{ampl}
{Ampleforth}. \url{https://www.ampleforth.org/} (2021)

\bibitem{bac}
{Basis Cash}. \url{https://basis.cash/} (2021)

\bibitem{dune}
{Dune Analytics}. \url{https://duneanalytics.com/} (2021)

\bibitem{frax}
{Frax}. \url{https://frax.finance/} (2021)

\bibitem{usdt}
{Tether}. \url{http://tether.to} (2021)

\bibitem{uniswap}
{Uniswap}. \url{http://uniswap.io} (2021)

\bibitem{alur1994theory}
Alur, R., Dill, D.L.: A theory of timed automata. Theoretical computer science
  \textbf{126}(2),  183--235 (1994)

\bibitem{ametrano2016hayek}
Ametrano, F.M.: Hayek money: The cryptocurrency price stability solution.
  Available at SSRN 2425270  (2016)

\bibitem{bullmann2019search}
Bullmann, D., Klemm, J., Pinna, A.: In search for stability in crypto-assets:
  are stablecoins the solution? ECB Occasional Paper (230) (2019)

\bibitem{caginalp2018dynamical}
Caginalp, C.: A dynamical systems approach to cryptocurrency stability. arXiv
  preprint arXiv:1805.03143  (2018)

\bibitem{caginalp2018opinion}
Caginalp, C., Caginalp, G.: Opinion: Valuation, liquidity price, and stability
  of cryptocurrencies. Proceedings of the National Academy of Sciences
  \textbf{115}(6),  1131--1134 (2018)

\bibitem{clark2019sok}
Clark, J., Demirag, D., Moosavi, S.: Sok: Demystifying stablecoins. Available
  at SSRN 3466371  (2019)

\bibitem{hileman2019state}
Hileman, G.: State of stablecoins (2019). Available at SSRN  (2019)

\bibitem{iwamura2019can}
Iwamura, M., Kitamura, Y., Matsumoto, T., Saito, K.: Can we stabilize the price
  of a cryptocurrency?: Understanding the design of bitcoin and its potential
  to compete with central bank money. Hitotsubashi Journal of Economics pp.
  41--60 (2019)

\bibitem{klages2020stablecoins}
Klages-Mundt, A., Harz, D., Gudgeon, L., Liu, J.Y., Minca, A.: Stablecoins 2.0:
  Economic foundations and risk-based models. In: Proceedings of the 2nd ACM
  Conference on Advances in Financial Technologies. pp. 59--79 (2020)

\bibitem{larsen1997uppaal}
Larsen, K.G., Pettersson, P., Yi, W.: Uppaal in a nutshell. International
  journal on software tools for technology transfer  \textbf{1}(1-2),  134--152
  (1997)

\bibitem{mita2019stablecoin}
Mita, M., Ito, K., Ohsawa, S., Tanaka, H.: What is stablecoin?: A survey on
  price stabilization mechanisms for decentralized payment systems. In: 2019
  8th International Congress on Advanced Applied Informatics (IIAI-AAI). pp.
  60--66. IEEE (2019)

\bibitem{moin2020sok}
Moin, A., Sekniqi, K., Sirer, E.G.: Sok: A classification framework for
  stablecoin designs. In: Financial Cryptography (2020)

\bibitem{nakamoto2019bitcoin}
Nakamoto, S.: Bitcoin: A peer-to-peer electronic cash system. Tech. rep.,
  Manubot (2019)

\bibitem{pernice2019monetary}
Pernice, I.G., Henningsen, S., Proskalovich, R., Florian, M., Elendner, H.,
  Scheuermann, B.: Monetary stabilization in cryptocurrencies--design
  approaches and open questions. In: 2019 Crypto Valley Conference on
  Blockchain Technology (CVCBT). pp. 47--59. IEEE (2019)

\bibitem{pnueli1977temporal}
Pnueli, A.: The temporal logic of programs. In: 18th Annual Symposium on
  Foundations of Computer Science (sfcs 1977). pp. 46--57. IEEE (1977)

\bibitem{saito2019make}
Saito, K., Iwamura, M.: How to make a digital currency on a blockchain stable.
  Future Generation Computer Systems  \textbf{100},  58--69 (2019)

\bibitem{sams2015note}
Sams, R.: A note on cryptocurrency stabilisation: Seigniorage shares. Brave New
  Coin pp.~1--8 (2015)

\bibitem{wood2014ethereum}
Wood, G., et~al.: Ethereum: A secure decentralised generalised transaction
  ledger. Ethereum project yellow paper  \textbf{151}(2014),  1--32 (2014)

\end{thebibliography}

\end{document}